# The MiniCLEAN Dark Matter Experiment

A. Hime for the MiniCLEAN Collaboration
*Physics Division, MS H803, Los Alamos National Laboratory, Los Alamos, NM 87545, USA*

The MiniCLEAN dark matter experiment will exploit a single-phase liquid-argon detector instrumented with photomultiplier tubes submerged in the cryogen with nearly $4\pi$ coverage of a 500 kg (150 kg) target (fiducial) mass. The high light yield and unique properties of the scintillation time-profile provide effective defense against radioactive background through pulse-shape discrimination and event-position reconstruction. The detector is designed also for a liquid-neon target that allows for an independent verification of signal and background and a test of the expected dependence of the WIMP-nucleus interaction rate.

## 1. Introduction

MiniCLEAN is the first major step in a family of detectors to search for Weakly Interacting Massive Particle (WIMP) dark matter using "single-phase" detectors of liquid argon (LAr) and liquid neon (LNe) as the target material. Our vision for CLEAN (Cryogenic Low Energy Astrophysics with Noble liquids) has evolved over the years. Originally conceived as a massive (order 100 tonne) detector of LNe, the unique properties of its scintillation light and ability to achieve extremely high target purity allow for a detector capable to detect, simultaneously, WIMP dark mater and low energy neutrinos from the Sun [1]. Our direction broadened with the realization that LAr can also serve as a suitable target in a single-phase detector owing to the unprecedented capability projected to discriminate electron recoils from nuclear recoils by comparing the relative populations of the singlet and triplet components of the scintillation time profile [2]. Dubbed DEAP for Dark matter Experiment with Argon and Pulse-shape discrimination, this powerful capability provides the necessary attack on $^{39}$Ar background inherent in a LAr target. MiniCLEAN, described here, and DEAP-3600 [3], share the same basic elements of the conceptually simple detector shown schematically in Figure 1, albeit with different technological and engineering approaches. Our overarching goal is a single-phase detector of WIMP dark matter with a target mass at the 40 to 120 tonne scale, the engineering design and scientific scope for which will be guided by the performance of the MiniCLEAN and DEAP-3600 experiments.

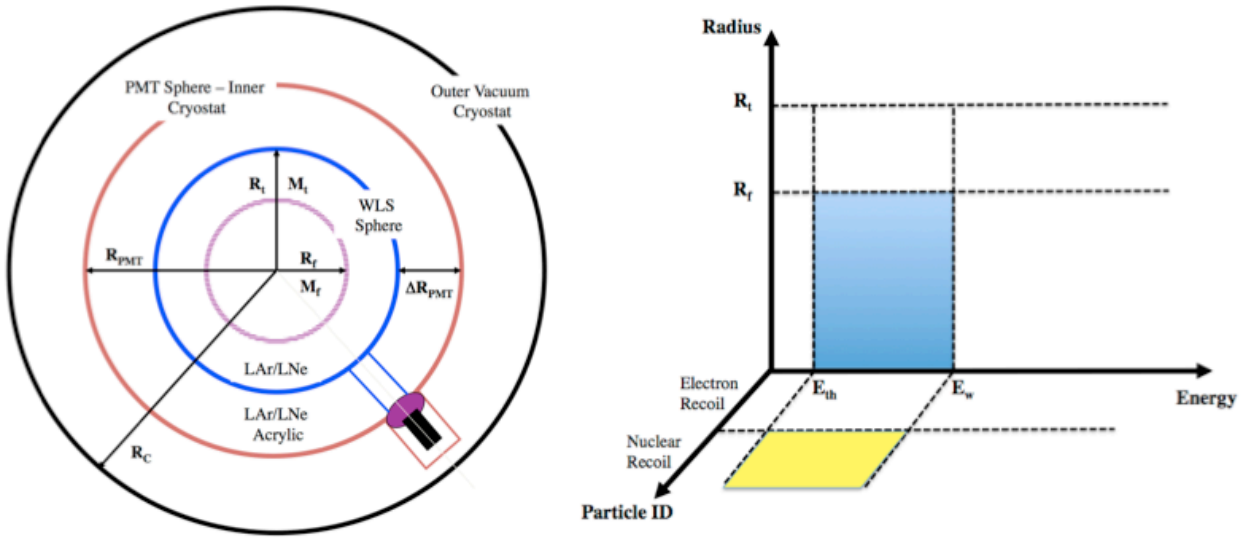

**Figure 1:** The left hand panel depicts the conceptually simple approach to a single-phase liquid scintillation detector wherein the cryogen is contained within a central target of radius $R_t$ and viewed in $4\pi$ by an array of photomultiplier tubes (PMTs). The inner surface of the target sphere is coated with a wavelength-shifting fluor that converts the extreme ultra-violet to the blue that is detected, via light guides, by the PMTs. An interstitial volume between the target and PMT sphere is filled with a combination of cryogen and/or acrylic to serve as shielding against fast neutrons produced, primarily, in the PMT glass. Event energy and position are reconstructed using the charge distribution collected by the PMT array, defining a fiducial volume with radius $R_f$ and energy window for analysis as depicted by the blue-shaded region in the right panel. Particle identification using pulse-shape discrimination (PSD) is used to separate background (electron recoils) from signal (nuclear recoils), depicted by the yellow-shaded region. It is the combination of PSD and fiducialization that provides the background-free energy window appropriate for a WIMP search.

With a target (fiducial) mass of 500 kg (150 kg), MiniCLEAN will exploit both LAr and LNe. DEAP-3600, with a target (fiducial) mass of 3600 kg (1000 kg), is dedicated to LAr. Table I provides a comparison of the salient differences between the MiniCLEAN and DEAP-3600 experiments in their engineering approach and defence against two important backgrounds, namely fast neutrons from the PMTs and surface contamination from radon daughters that will plate out on the inner wavelength-shifting surface. MiniCLEAN takes a modular approach where the 92 optical cassettes will be assembled in a radon-free glove box and then transported and inserted into the Inner Vessel under vacuum. The combination of acrylic and interstitial cryogen serves as neutron shielding with the PMTs submerged "cold" in the cryogen. DEAP-3600 will use a monolithic acrylic vessel to contain the cryogen with the PMTs operating "warm" and shielded by 50 cm of acrylic. Radon daughters will be removed *in situ* using a mechanical device to resurface the inside of the vessel.

**Table I:** Technical Specifications of the MiniCLEAN and DEAP-3600 Detectors.

|  | **MiniCLEAN** | **DEAP-3600** |
|---|---|---|
| **Target Capability** | LAr & LNe | LAr |
| **Target Radius (cm)** | 45 cm | 85 cm |
| **Target Mass (kg)** | 500 | 3600 |
| **Fiducial Mass (kg)** | 150 | 1000 |
| **Light Collection** | 92 Modular Optical Cassettes with PMTs Submerged "Cold" | 266 "Warm" PMTs outside of Cryogen |
| **Cryogenic Containment** | Code Stamped Stainless Steel Pressure Vessel | Monolithic Acrylic Vessel |
| **Neutron Shielding** | 10 cm Acrylic + 20 cm Cryogen | 50 cm Acrylic |
| **Surface Radon Mitigation** | Modular Cassettes Assembled in Vacuum | In Situ Resurfacing of inner Acrylic Vessel Surface |
| **Process Systems** | Pulse-Tube Refrigerators with Heat Exchange | LN-Cooled Thermal Siphon |

## 2. The MiniCLEAN Detector

A conceptual design of the MiniCLEAN detector is provided in Figure 2 below. The central detector is composed of three major elements, an Inner Vessel (IV) that contains the liquid cryogen, an array of optical cassettes that are inserted into the IV and define the inner target region, and an Outer Vessel (OV) that provides secondary containment and the necessary thermal insulation of the inner cryostat. Progress on the fabrication of the IV and OV is shown in Figure 3. The entire detector will be shielded in a water tank with an active muon veto and operated 6800 feet underground in the Cube Hall at SNOLAB (Figure 4).

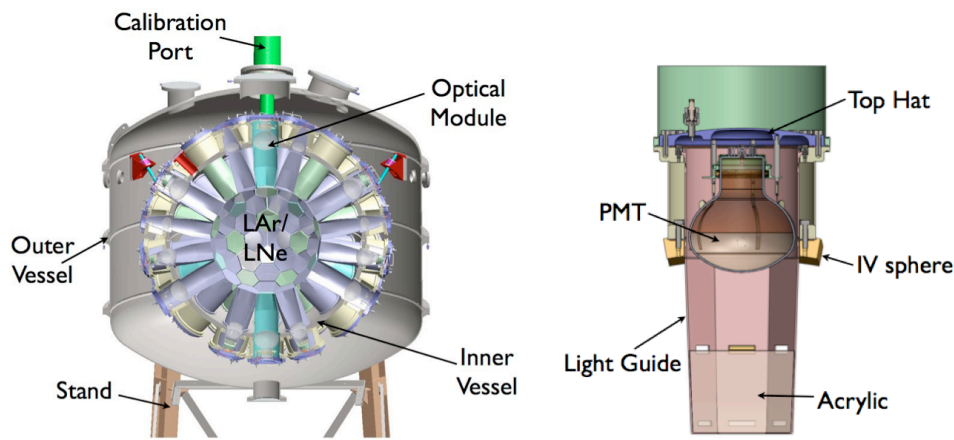

**Figure 2:** Conceptual design of the MiniCLEAN central detector (left) with its $4\pi$ target viewed by 92 optical cassettes. The optical cassettes (right) consist of an acrylic plug, the front surface of which is coated with a wavelength-shifting fluor (TPB), and light guide leading to the PMT. The array of optical modules will be assembled in a radon-free environment and inserted under vacuum into the Inner Vessel. The central detector consisting of the Inner Vessel and optical cassettes is enclosed and thermally insulated under vacuum by the Outer Vessel.

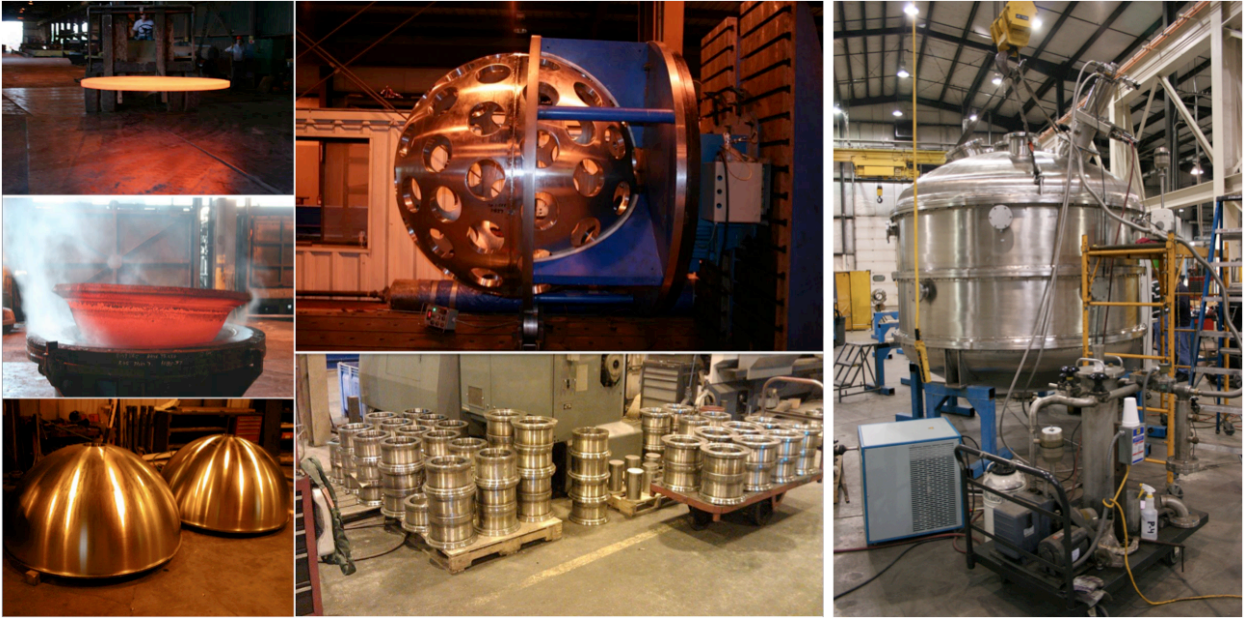

**Figure 3:** Shown in the three small photographs to the left are the early stages of forming the two hemispheres that make up the Inner Vessel (IV). The IV is being fabricated by Winchester Precision Technologies in Winchester, NH as an ASME Section VIII, Division-1 pressure vessel. The central-top picture shows a recent photograph of the IV with its 92 ports and the two hemispheres successfully welded together. A subset of the spools is shown at the bottom center. These spools will serve as the ports for the optical cassettes that will be bolted to the IV via conflat-seal. The Outer Vessel (OV) is shown at the far right and has been fabricated at PHPK Technologies in Columbus, OH. Also a code-stamped vessel, it is made up of four sections to allow transportation underground at SNOLAB. Both the IV (64 inches in diameter) and OV (104 inches in diameter and 106 inches high) are manufactured from low-radioactivity stainless steel and were designed by the engineering team at LANL.

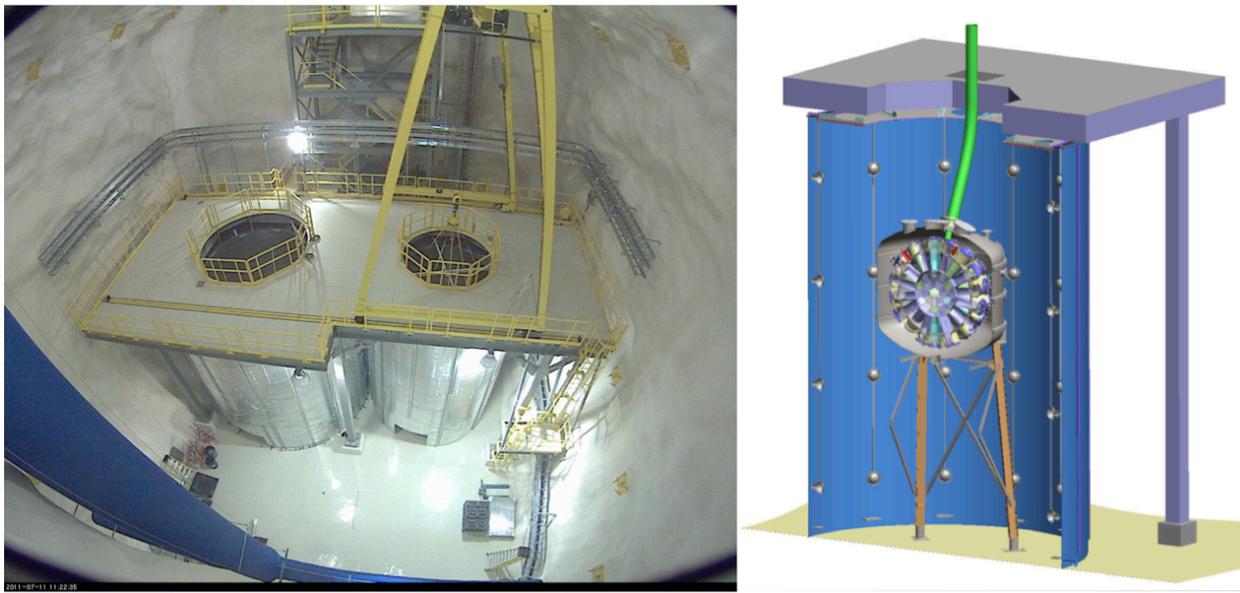

**Figure 4:** The SNOLAB Cube Hall (left) in July 2011 with the deck infrastructure and overhead 10-ton gantry crane installed. The water shield tanks for DEAP-3600 and MiniCLEAN have been constructed and are visible below the deck. The Cube Hall measures 50x50x60 feet and is presently operating as a Class-2000 clean room. The MiniCLEAN detector is shown conceptually on the right in its active water shield and supported by a stand that is being engineered to withstand seismic forces that might occur in the active nickel mine. The shield tank for MiniCLEAN is 18 feet in diameter and 25 feet high and equipped with a muon veto composed of 48 PMTs.

One of the great advantages of the MiniCLEAN design is that it is simple to model the detector's response to our backgrounds and the signals of interest. The techniques we use are nearly identical to those used in the large successful neutrino experiments, like SNO and Super-Kamiokande, in which the detector response depends primarily on bulk optical properties of the materials (attenuation and scattering lengths, reflectivities, and PMT response). The optics of the detector are simulated within the particle propagation framework of GEANT4 with material properties extracted from the literature along with measurements we have made in the laboratory for specific detector elements such as the tetraphenyl butadiene (TPB) wavelength shifter [4], PMTs and acrylic to be used in the optical cassettes. Valuable data have been acquired from our small prototype detectors that demonstrate the basic concepts for a single-phase detector and that serve to further benchmark and constrain our simulations for MiniCLEAN. The prototype detectors operated at Yale University with "cold" PMTs and using both LAr [5,6] and LNe [7] serve as the logical benchmark for MiniCLEAN. Results from measurements in microCLEAN of the effective scintillation yield for nuclear recoils and of PSD in LAr are shown in Figure 5. The DEAP-1 detector designed as a low-background experiment and operating at SNOLAB [3,8], has been instrumental in pushing the limits on PSD in single-phase LAr, the implications of which are discussed further in section 3.

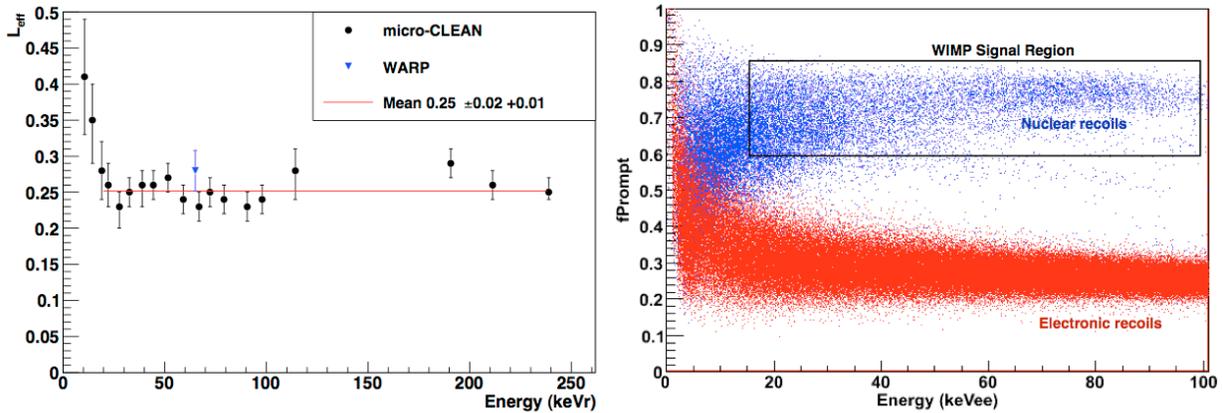

**Figure 5:** Results from the microCLEAN detector operated with LAr. The left panel shows measurements (ref.[5]) of the scintillation light yield for nuclear recoils relative to electrons ($L_{eff}$). The right panel shows the PSD between nuclear recoils and electron recoils as a function of energy (ref.[6]). A light yield of ~6 pe/keV has been demonstrated in the microCLEAN experiment.

## 3.  A Background Model for the MiniCLEAN Detector

Mitigation of radioactive background poses a significant challenge to the design of the experiment, the choice of construction materials and assembly procedures. Gamma rays and fast neutrons from radioactivity in the cavern rock are effectively attenuated by the 150 cm water shield and SNOLAB depth is more than sufficient for screening out cosmic ray muon activity and muon-induced fast neutrons, in particular [9]. The dominant backgrounds in MiniCLEAN arise from $^{39}$Ar in the target, radon contamination of the inner WLS surface, and fast neutrons from radioactivity in the PMT glass. Additional backgrounds from detector materials beyond the PMT sphere contribute ~10 to 15% of the total background budget and we focus on the three main contributors below.

## 3.1.  Pulse-Shape Discrimination and $^{39}$Ar Background

$^{39}$Ar is a beta emitter with endpoint energy of 565 keV and a half-life of 269 years. It is present in natural argon at ~8 parts in $10^{16}$, yielding the intimidating rate of ~1 Bq/kg. A rejection factor of a few parts in $10^9$ for low-energy electrons is required for MiniCLEAN in order to reach a background rate of < 1 event per year. Results from the DEAP-1 experiment [3,8] are now approaching these levels and, while statistically limited, project discrimination beyond the $10^{-9}$ level as advertised in ref.[2]. In order to project the limits obtained in DEAP-1 to the MiniCLEAN detector a detailed simulation is required that takes account of the multi-channel configuration, optics and light propagation, event reconstruction, and PMT response. We have done so for the MiniCLEAN detector, the results of which are shown in Figure 6. When the upper bound from DEAP-1 is combined with our simulation of the MiniCLEAN detector response, the integral leakage of $^{39}$Ar events above an energy threshold of 20 keV$_{ee}$ (120 pe) is less than 2 events per year in our 150 kg fiducial volume. This upper limit is projected to be at least an order of magnitude smaller (< 0.2 events / 150 kg / year) when sacrificing an acceptance of 50% in the nuclear recoil region of interest for $f_{prompt}$.

Segmentation of the MiniCLEAN detector into a large number of PMT channels improves the quality of information significantly when compared to the smaller prototype detectors with just two PMTs. Instead of performing particle identification by integrating the summed waveforms over two time windows, as is done to compute the $f_{prompt}$ statistic, we can instead identify the time of arrival of individual photoelectrons at each PMT. We can use the time of each photoelectron to produce a powerful particle identification statistic we call $L_{recoil}$. It is a likelihood ratio computed using the observed time of arrival for all of the photoelectrons in the event. The likelihood for the observed times is computed under the hypothesis of a single nuclear recoil (i.e. a WIMP candidate) and divided by the likelihood under the hypothesis that the event is an electron recoil. By computing $L_{recoil}$ after photoelectron counting, we have effectively removed the effects of the broad PMT charge distribution, producing a test statistic with much less variance than $f_{prompt}$. As Figure 6 shows, this reduction in variance significantly improves the rejection of electron recoils with $L_{recoil}$ compared to $f_{prompt}$.

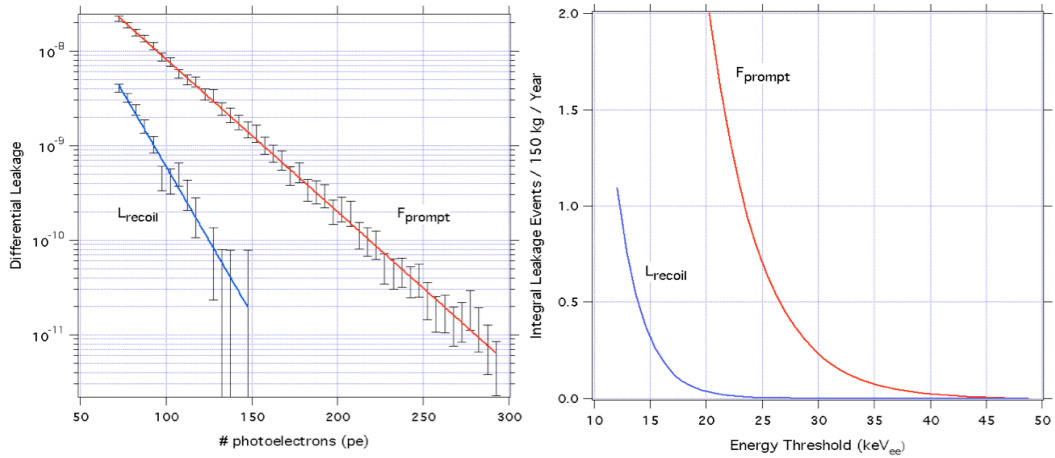

**Figure 6:** The left panel shows the differential leakage of electron events (points with statistical uncertainties) as simulated in our MiniCLEAN Monte Carlo program. The solid curves are exponential fits to the data points for the $F_{prompt}$ and $L_{recoil}$ statistics and constrained by the upper limits obtained in the DEAP-1 experiment. The right panel shows the integral leakage of $^{39}$Ar events as a function of energy threshold and into a MiniCLEAN fiducial volume of 150 kg.

## 3.2. Surface Radon and Its Progenies

Radon daughters and their progenies that plate out on the wavelength shifter surface viewing the target volume present a significant hazard for a sensitive WIMP search (see Figure 7). Decay rates of ~1/m$^2$/day would yield a total of ~500 recoil nuclei per year directed into the target. This level of surface activity would accumulate in less than half a day in a typical room with a radon concentration of about 10 Bq/m$^3$[10]. With radon-contamination under control at assembly, the background rate can be reduced by about a factor of 1000 within the fiducial radius of 29.5 cm. We predict it is possible to reduce this 0.5 events/year/150 kg by another order of magnitude by exploiting the fact that the alpha particle that accompanies the decay scintillates in the TPB [11].

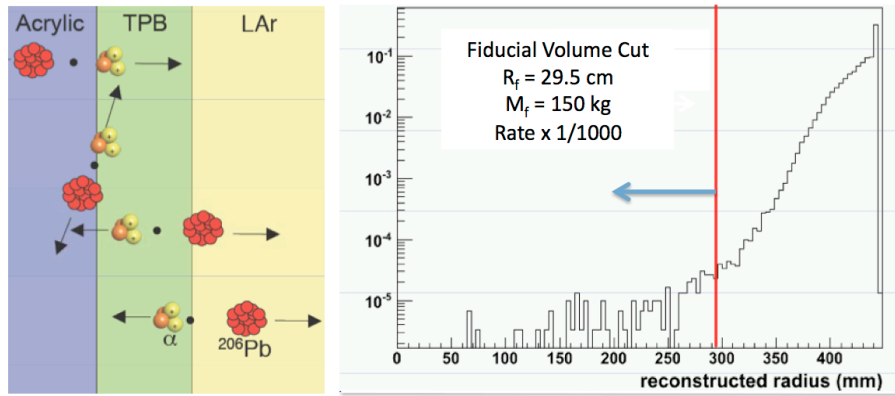

**Figure 7:** Daughter nuclei from surface activity (left panel) can produce recoil events that are a background to a WIMP signal. They are effectively reduced in MiniCLEAN by position reconstruction and fiducialization (right panel).

## 3.3. Fast Neutrons

Fast neutrons reaching the central target are dominated by ($\alpha$, n) reactions in the PMT glass. Roughly 90% (10%) of these neutrons are produced in the $B_2O_3$ ($SiO_2$) constituting the bulk of the glass. Based on radioassay of the uranium and thorium content of the glass, we predict that 42,000 neutrons will be produced from MiniCLEAN's PMT array in one year. These neutrons are reduced to about 1073 per year due to the 30 cm of shielding (10 cm of acrylic + 20 cm cryogen) between the PMT sphere and target volume. Fiducialization and "tagging" multiple interactions reduces the fast neutron background to a tolerable level as shown in Table II. Since our neutron background is dominated by radioactivity in the borated silica glass of the PMTs we are developing an encapsulated source that mocks up this source as closely as possible. Data from this source and a pulsed D-D generator will be used to validate our simulation's distribution of neutrons in reconstructed radius, energy and pulse shape, permitting the observed PMT-neutron-induced events outside the region of interest to be extrapolated into the WIMP analysis window.

Table II: Fast Neutron Backgrounds from the PMT Array in MiniCLEAN-LAr

| Event Selection Criteria | Number of ($\alpha$, n) Events |
|---|---|
| None – Generated at PMT Array | 42000 |
| Number Reaching Target with R < 44 cm & E < 25 keV$_{ee}$ | 1073 |
| Reconstructed with E > 12.5 keV$_{ee}$ | 158.3 ± 1.5 |
| & within Fiducial R < 29.5 cm | 30.4 ± 0.7 |
| & with $F_{prompt}$ > 0.7 | 0.9 ± 0.1 |
| Reconstructed with E > 20 keV$_{ee}$ | 49.6 ± 0.8 |
| & within Fiducial R < 29.5 cm | 10.2 ± 0.4 |
| & with $F_{prompt}$ > 0.7 | 0.2 ± 0.02 |

## 4. Exploiting $^{39}$Ar as a Calibration Device

The *in situ* decay rate (~1 Hz/kg) of $^{39}$Ar in natural argon is such that one would accumulate ~$10^6$ events/10 kg/day. Consequently, the general state of the detector (single p.e. response and PMT performance, electronic noise, light yield, target purity) and its temporal stability can be monitored and calibrated on a daily basis. The $^{39}$Ar is uniformly distributed throughout the target volume and thus can be used to calibrate and monitor position reconstruction as a function of energy by dividing the sample into (10 kg) spherical shells. As $^{39}$Ar is a unique first-forbidden beta decay, its spectral shape is well defined and can be used to calibrate and monitor the absolute energy scale and resolution of the detector (see Figure 8).

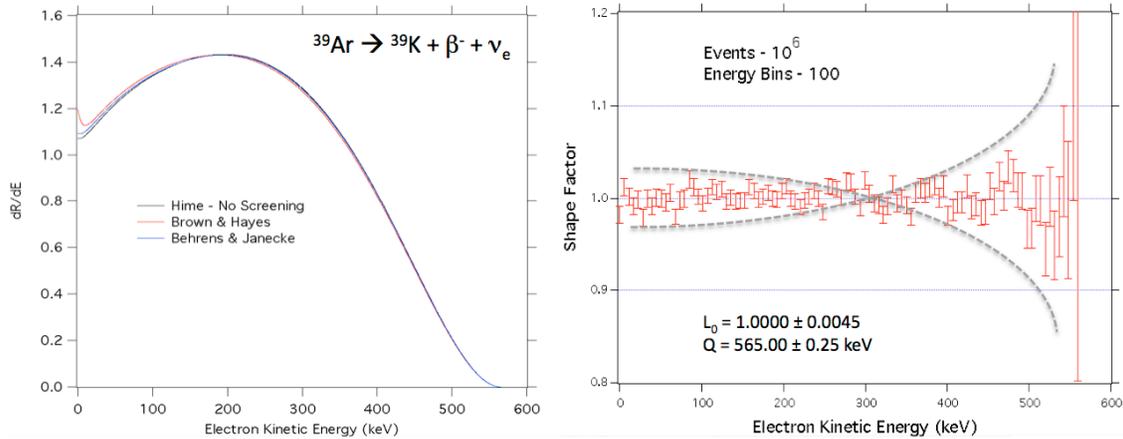

**Figure 8:** The left panel shows the differential energy spectrum for the unique first-forbidden decay of 39Ar. Calculations have been made for three different methods of handling the Coulomb corrections. The right panel shows the residual shape factor resulting from a simulated fit to $10^6$ beta decay events where the endpoint energy is allowed to float and is determined to better than 0.25 keV (compared with the tabulated value of 565 ± 5 keV). The dashed-grey curves show the distortion in the residual shape when the effective energy scale ($L_0$) is varied by ± 0.5%. These preliminary results indicate that the energy scale can be determined *in situ* on a daily basis (and as a function of reconstructed position) to a precision of ~1% when consideration is provided for uncertainties in the Coulomb corrections (ref.[12]).

## 5. Scientific Reach and Scalability

Once commissioned, we plan to begin a scientific run with LAr with the data blinded in a WIMP search region of interest (ROI) based upon the observable degrees of freedom, namely event energy, position, and particle-ID. The ROI will be defined based upon simulations that are benchmarked against calibration data. Our baseline for projecting the sensitivity of the MiniCLEAN experiment is presently an energy interval of 12.5 to 25 keV$_{ee}$ (50 to 100 keV$_r$), a fiducial mass of 150 kg defined by events reconstructing within 29.5 cm, and nuclear recoil particle-ID with $f_{prompt} > 0.7$. Sensitivity contours for the WIMP-nucleon Cross Section as a function of the WIMP mass are shown in Figure 9 along with a flowchart describing the utility of implementing a $^{39}$Ar spike at the end of the LAr run and the benefits of target exchange with LNe.

A 2-year run with natural argon in MiniCLEAN can establish sensitivity at the $10^{-45}$ cm$^2$ level for the larger WIMP masses motivated by SUSY models. The result would be competitive with the liquid xenon programs of XENON-100 [13] and LUX [14], and would establish the database for comparison to the performance of the dual-phase LAr programs of DarkSide [15] and WArP [16] as recommended by DMSAG [17] and PASAG [18]. At the tonne scale, DEAP-3600 can extend the reach to $10^{-46}$ cm$^2$ with natural argon. We have also made some preliminary projections for a full-scale CLEAN detector using a 40-tonne (10-tonne) target (fiducial) mass as a baseline. A 10-tonne LNe detector has sensitivity comparable to that of a 1-tonne LAr experiment and has the advantage of being free of internal radioactivity like $^{39}$Ar. It is at this scale that the benefits of argon depleted in $^{39}$Ar begins to become apparent[1].

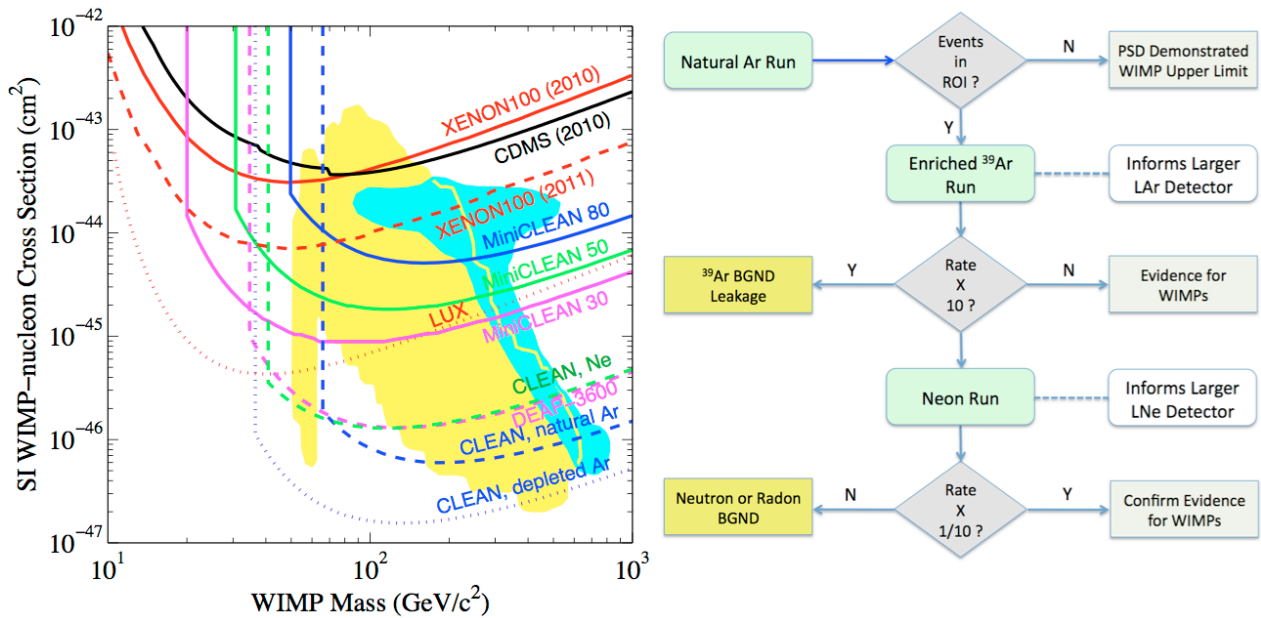

**Figure 9:** The left panel shows our projected zero-background sensitivity (90% C.L.) on the WIMP-nucleon spin-independent cross section for MiniCLEAN assuming threshold energies of 30 keVr, 50 keVr and 80keVr. Prior upper limits obtained from CDMS and XENON100, along with projections for LUX are shown for comparison. Projections for a 1-tonne fiducial DEAP-3600 experiment and a 10-tonne fiducial CLEAN experiment with LNe and LAr are also shown. For all curves, the standard halo of Donato *et al* with escape velocity=544 km/s is assumed. The shaded regions indicate allowed parameter space calculated from Constrained Minimal SUSY Models by Baltz and Gondolo (yellow) and Trotta *et al* (cyan). The right panel depicts the logic behind a $^{39}$Ar spike and target exchange with LNe. A spike of $^{39}$Ar in MiniCLEAN will allow a precise test of PSD rejection efficiency. The added ability to exchange the target with LNe will allow a test of signal against a misinterpreted background such as fast neutrons or radon decay. In this way, MiniCLEAN serves to inform the design and capability of larger scale, single-phase detectors and, effectively, performs as a "Beam-On, Beam-Off" detector of WIMP dark matter.

---

[1] This should be contrasted with a dual-phase LAr detector that requires argon depleted of $^{39}$Ar starting at order 100 kg target masses owing to the long electron drift time (~1 to 3 ms) in comparison to the triplet-lifetime (1.5 μs) exploited in the single-phase approach without electric fields.

## 6. Summary & Outlook

Construction of the MiniCLEAN detector is well under way and marks a major step toward the use of single-phase LAr in the search for WIMP dark matter. The high light yield projected and $4\pi$ coverage allows the necessary attack against radioactive background through pulse-shape discrimination and fiducialization. The ability to implement a radioactive $^{39}$Ar spike into the active volume of the detector and to exchange the LAr target with LNe is unique to MiniCLEAN, offering a novel way to exercise potential signals and backgrounds and the means to inform the design of larger scale detectors. The conceptual simplicity of the design provides a straightforward path to detectors in excess of 10 tonnes.

## Acknowledgments

The author is grateful to the organizers of the DPF-2011 conference and for the opportunity to speak on behalf of the MiniCLEAN collaboration. The MiniCLEAN experiment is supported by the LDRD Program at Los Alamos National Laboratory, the US Department of Energy Office of Science, the National Science Foundation, and the Packard Foundation.

## References

[1] D.N. McKinsey and K.J. Coakley, Astropart. Phys. **22**, 355 (2005); M.G. Boulay, A. Hime and J. Lidgard, arXiv:nucl-ex/0410025 (2004); D.N. McKinsey and J.M. Doyle, J. Low Temp. Phys. **118**, 153 (2000).
[2] M.G. Boulay and A. Hime, Astropart. Phys. **25**, 179 (2006).
[3] M.G. Boulay, TAUP, Munich, Germany, Sept.8 (2011), http://taup2011.mpp.mpg.de/; M.G. Boulay and B. Cai, J. Phys.: Conf. Ser. **136**, 042081 (2008).
[4] V.M. Gehman, *et. al.,* arXiv:1104.3259 (2011).
[5] LAr quenching factor from microCLEAN …
[6] W.H. Lippincott, *et. al.,* Phys. Rev. C**78**, 035801 (2008).
[7] W.H. Lippincott, *et. al., work in progress*; J.A. Nikkel, R. Hasty, W.H. Lippincott and D.N. McKinsey, Astropart. Phys. **29**, 161 (2008).
[8] M.G. Boulay, *et. al.,* arXiv:0904.2930 (2009).
[9] D.-M. Mei and A. Hime, Phys. Rev. D**73**, 053004 (2006).
[10] V.E. Guiseppe, S.R. Elliott, A. Hime, K. Rielage and S. Westerdale*,* AIP Conf. proc. **1338**, 95 (2011).
[11] T. Pollman, M.G. Boulay and M. Kuzniak, arXiv:1011.1012 (2010).
[12] L.S. Brown, A.C. Hayes and A. Hime, *The $^{39}$Ar spectrum is uniquely and analytically defined down to Coulomb corrections which must include screening of the nuclear charge due to the orbital electrons. These corrections become important at low energies (< 10 keV) where the beta kinetic energy is comparable to the screening potential.*
[13] E. Aprile *et. al.,* XENON-100 Collaboration, arXiv:1104.2549 (2011).
[14] J. Chapman, for the LUX Collaboration, *in these proceedings*.
[15] A. Wright, for the DarkSide Collaboration, *in these proceedings*.
[16] WArP Collaboration, J. Phys.: Conf. Ser. **203**, 012006 (2006); Astropart. Phys. **28**, 495 (2008).
[17] The Dark Matter Scientific Advisory Group, A Joint Sub-panel of HEPAP and AAAC Report on the Direct Detection and Study of Dark Matter (July 5, 2007), www.science.doe.gov/hep/DMSAGReportJuly18,2007.pdf.
[18] The Particle Astrophysics Scientific Assessment Group, Report of the HEPAP PASAG (October 23, 2009), www.er.doe.gov/hep/files/pdfs/PASAG_Report.pdf.